\documentclass[aps,pre,epsfig,preprint,showpacs]{revtex4}
\usepackage{graphicx,bm,epsfig}
\begin{document}
\newcommand{\lb} {\makebox{{\LARGE [}}}
\newcommand{\rb} {\makebox{{\LARGE ]}}}
\newcommand{\gae}
{\,\hbox{\lower0.5ex\hbox{$\sim$}\llap{\raise0.5ex\hbox{$>$}}}\,}
\newcommand{\lae}
{\,\hbox{\lower0.5ex\hbox{$\sim$}\llap{\raise0.5ex\hbox{$<$}}}\,}
\title{Critical line of an $n$-component cubic model}
\author{Wenan Guo$^{1,5}$\footnote{E-mail:waguo@bnu.edu.cn}, 
Xiaofeng Qian$^{2}$,
Henk W. J. Bl\"ote$^{3,2}$
and F. Y. Wu$^{4}$
}
\affiliation{$^{1}$ Physics Department, Beijing Normal
University, Beijing 100875, P. R. China \\
$^{2}$ Instituut Lorentz, Universiteit Leiden,
Niels Bohrweg 2, Postbus 9506, 2300 RA Leiden, The Netherlands \\
$^{3}$Faculty of Applied Sciences, Delft University of
Technology,  P.O. Box 5046, 2600 GA Delft, The Netherlands \\
$^{4}$ Department of Physics, Northeastern University,
Boston, Massachusetts 02115, USA \\
$^{5}$ The Abdus Salam International Centre for Theoretical
Physics, Trieste, Italy }
\date{\today}
\begin{abstract}
We consider a special case of the $n$-component cubic model on the
square lattice, for which an expansion exists in Ising-like graphs.
We construct a transfer matrix and perform a finite-size-scaling
analysis to determine the critical points for several values of $n$.
Furthermore we determine several universal quantities,
including three critical exponents.
For $n<2$, these results agree well with the theoretical predictions
for the critical O($n$) branch.
This model is also a special case of the ($N_\alpha,N_\beta$)
model of Domany and Riedel. It appears that the self-dual plane of
the latter model contains the exactly known critical points of the
$n=1$ and 2 cubic models. For this reason we have checked whether this
is also the case for  $1<n<2$. However, this possibility is excluded
by our numerical results.

\end{abstract}
\maketitle

\section {Introduction}
\label{intro}
The $n$-component cubic model can be defined in terms of vector spins
that are restricted to lie along one of $n$ Cartesian axes, but are free
to assume the positive or negative direction. Only one vector component
is nonzero; it is normalized to be $\pm 1$. 
The model can be described by the reduced Hamiltonian
\begin{equation}
{\mathcal H}/k_{\rm B}T = - \sum_{\langle{ij}\rangle}
[K \vec{s_i} \cdot \vec{s_j} + M  (\vec{s_i} \cdot \vec{s_j})^2]\,,
\label{Hcub}
\end{equation}
where the index $i$ of the cubic spin $\vec{s}_i$ refers to the sites of
the square lattice.  The sum on $\langle{ij}\rangle$ stands for all
nearest-neighbor pairs.
This model obviously
combines Potts degrees of freedom (the choice of the Cartesian axis, 
which is subject to permutation symmetry) with Ising degrees of freedom
which specify the sign of the nonzero component.
This  is more explicit in the following form of the Hamiltonian
\begin{equation}
{\mathcal H}/k_{\rm B}T = - \sum_{\langle{ij}\rangle}
(K s_i s_j + M ) \delta_{\tau_i \tau_j} \,, 
\label{Hcub1}
\end{equation}
where we represent the sign of the nonzero component of $\vec{s_i}$ by
$s_i=\pm1$ and its Cartesian axis number by $\tau_i=1,2,\cdots,n$.
The corresponding bond weight $w_{ij}$ can be rewritten as
\begin{eqnarray}
w_{ij}&=&\exp[(Ks_i s_j + M ) \delta_{\tau_i \tau_j}] \nonumber \\
 &=&1+\delta_{\tau_i \tau_j} \{\exp(K s_i s_j+M) -1\} \nonumber \\
 &=&  1+ \delta_{\tau_i \tau_j} n \{v+ x s_i s_j \}   \nonumber \\
 &=&\sum_{b_{ij}=0}^1[\delta_{\tau_i \tau_j}n\{v+xs_is_j\}]^{b_{ij}} \,.
\end{eqnarray}
In the third line we have used the definitions
\begin{equation}
v \equiv \frac{e^M \cosh K -1}{n} {\rm ~~and ~~} 
x \equiv \frac{e^M \sinh K }{n} \,,
\label{defab}
\end{equation}
and in the fourth line we have introduced bond variables $b_{ij}=0$ or 1,
and the summand is subject to the rule $0^0=1$.
Thus the partition sum assumes the form
\begin{equation}
Z_{{\rm cub}} =  \sum_{\{s\},\{\tau\}} \prod_{\langle{ij}\rangle}
\sum_{b_{ij}=0}^1 [\delta_{\tau_i \tau_j} n \{v+ x s_i s_j \}]^{b_{ij}} \,.
\label{Zcub2}
\end{equation}
Application of the Kasteleyn-Fortuin mapping \cite{KF}
involves execution of the sum on the Potts-like variables $\{\tau\}$.
This leads to
\begin{equation}
Z_{{\rm cub}} = n^N \sum_{\{s\}} \lb \prod_{\langle{ij}\rangle} 
 \sum_{b_{ij}=0}^1 \{v+ x s_i s_j \}^{b_{ij}}\rb n^{n_l} \,,
\label{Zcub3}
\end{equation}
where $N$ is the number of sites of the lattice. Note that each bond
(by which we mean a bond variable $b_{ij}=1$) contributes, through the
Kronecker $\delta$, also a factor $1/n$, unless it closes a loop.
The latter condition  is accounted for by the factor $n^{n_l}$, where
$n_l$ is the number of loops formed by the bond variables.
Each configuration of bond variables $b_{ij}$ defines a graph on the
square lattice covering those and only those edges for which $b_{ij}=1$.
For the special case 
\begin{equation}
\cosh K = e^{-M}  {\rm ~~or~~} v=0  \,,
\label{specc}
\end{equation}
the partition sum reduces, after execution of the sum on $\{s\}$, to
\begin{equation}
Z_{{\rm cub}} = (2n)^{N} \sum_{\{b\}} x^{n_b} n^{n_l} \,,
\label{Zcub4}
\end{equation}
where 
the sum on $\{b\}$ contains only `even' graphs in which every site
is connected to an even number of bonds $b_{ij}=1$. The odd graphs,
while included in the sums on the $b_{ij}$ in Eq.~(\ref{Zcub3}), do
not survive the sum on $\{s\}$. The number of bonds in the
graph  $\{b\}$ is denoted 
$n_b \equiv \sum_{\langle{ij}\rangle}b_{ij}$.

The cubic loop model described by Eq.~(\ref{Zcub4}), is subject to
the restriction $|xn| \leq 1$ because of  Eq.~(\ref{specc}).
This model, and the model defined by Eq.~(\ref{Hcub}) with $M=0$,
(both on the square lattice) have already been investigated by a
transfer-matrix method \cite{BNcub}. That work included a determination
of the temperature exponent.  The results for $n<2$ were in agreement
with a conjecture of Cardy and Hamber \cite{CH} and the analysis of
Nienhuis \cite{N}. However, the latter work applies to the 
$n$-component cubic model on the honeycomb lattice, of which the graph
expansion reduces to that of the O($n$) model. For other lattices,
no such direct correspondence exists, and the relevance of the results
of Ref.~\onlinecite{N} for the cubic model on the square lattice thus
needs further justification. This is provided by renormalization
arguments that predict that cubic anisotropy is irrelevant
\cite{NienhuisD&L} for $n<2$. However, for $n=2$ it is marginal and
indeed the temperature exponent does not agree with the O(2)
model \cite{BNcub}.

The work of Ref.~\onlinecite{BNcub} did not include results for the
interval $1<n<2$, where the emphasis lies of the present work.
Furthermore, the present analysis will purportedly yield more
information about universal quantities.

The outline of this paper is as follows. Section \ref{sdplane} discusses
a duality transformation in a three-dimensional parameter space containing
the cubic model Eq.~(\ref{Zcub4}). This transformation suggests a possible
form of the critical line of the cubic model. However, transfer-matrix
calculations, defined in Sec.~\ref{trmat}, yield numerical results listed
in Sec.~\ref{crline}, which show that this form is not applicable.
The universal properties of the cubic model are investigated in 
Sec.~\ref{uniprop}. The paper concludes with a short discussion in
Sec.~\ref{disc}.

\section {Self-dual plane of the ($N_\alpha,N_\beta$) model }
\label{sdplane}

The partition function of the ($N_\alpha,N_\beta$) model \cite{DR} 
is defined as 
\begin{equation}
Z_{(m,n)} (e^{K_0},e^{K_1},e^{K_2},e^{K_3})\equiv
\sum_{\sigma=1}^m \sum_{\tau=1}^n \prod_{\rm edges}
B(\sigma,\tau|\sigma',\tau') \,,
\label{psummn}
\end{equation}
where $m=N_\alpha$, $n=N_\beta$, and 
\begin{displaymath}
B(\sigma,\tau|\sigma',\tau') \equiv
\exp[K_0 \delta_{\sigma \sigma'} \delta_{\tau \tau'}+
     K_1 (1-\delta_{\sigma \sigma'}) \delta_{\tau \tau'}+
\end{displaymath}
\begin{equation}
\mbox{\hspace{10mm}}
     K_2 \delta_{\sigma \sigma'} (1-\delta_{\tau \tau'})+
     K_3 (1-\delta_{\sigma \sigma'}) (1-\delta_{\tau \tau'})]\,.
\label{Bfacmn}
\end{equation}
Just as the cubic model, the model can be viewed as having
two Potts spins $\sigma$ and $\tau$  on each lattice site, with
allowed values $\sigma=1,2,\cdots,m$ and $\tau=1,2,\cdots,n$. These
spins interact with nearest-neighbor couplings according to the
Boltzmann weights (\ref{Bfacmn}).
Note that these weights are elements of the $(mn) \times (mn)$
matrix
\begin{equation}
{\bf B} = e^{K_0} I_m \otimes I_n + e^{K_1} J_m \otimes I_n +
          e^{K_2} I_m \otimes J_n + e^{K_3} J_m \otimes J_n \,, 
\end{equation}
where $I_n$ and $J_n$ are $n \times n$ matrices
\begin{equation}
I_n=\left(\begin{array}{cccc}
1 & 0 &\cdots& 0\\
0 & 1 &\cdots& 0\\
\vdots &\vdots &\ddots& \vdots\\
0 & 0 &\cdots& 1\\
\end{array}\right) \,,
\qquad
J_n=\left(\begin{array}{cccc}
0 & 1 &\cdots& 1\\
1 & 0 &\cdots& 1\\
\vdots &\vdots &\ddots& \vdots\\
1 & 1 &\cdots& 0\\
\end{array}\right) \,.
\end{equation}

We consider a model on a planar lattice with open boundary conditions.
The partition sum (\ref{psummn}) possesses the duality
relation \cite{WW}
\begin{displaymath}
Z_{(m,n)} (e^{K_0},e^{K_1},e^{K_2},e^{K_3})=  \mbox{\hspace{30mm}}
\end{displaymath}
\vspace{-5mm}
\begin{equation}
\mbox{\hspace{10mm}}
(mn)^{1-N_D}
Z^{(D)}_{(m,n)} (e^{K_0^\ast},e^{K_1^\ast},e^{K_2^\ast},e^{K_3^\ast}) \,,
\label{psdual}
\end{equation}
where $N_D$ is the number of sites of the dual lattice, and $Z^{(D)}$
the dual partition function, with Boltzmann weights also given by
Eq.~(\ref{Bfacmn}) but with the $K_i$ replaced by the dual weights
$K_i^\ast$. The latter weights were shown \cite{WW} to be equal to the
eigenvalues of the matrix ${\bf B}$. For the $(m,n)$ model, these
are \cite{DR}
\begin{equation}
\left(\begin{array}{c}
e^{K^*_0}\\ e^{K^*_1}\\ e^{K^*_2}\\ e^{K^*_3}\\
\end{array}
\right)
=\left(\begin{array}{cccc}
1 &m-1 & n-1 & (m-1)(n-1)\\
1 & -1 & n-1 & -(n-1)\\
1 &m-1 &  -1 & -(m-1)\\
1 & -1 &  -1 & 1\\
\end{array}
\right) 
\left(\begin{array}{c}
e^{K_0}\\
e^{K_1}\\
e^{K_2}\\
e^{K_3}\\
\end{array}
\right) \,.
\label{dual}
\end{equation}
Note that (\ref{dual}) implies $e^{(K_i^*)^*}=(mn)e^{K_i}$, and hence
$(Z^{(D)})^{(D)}=Z$, because the number of lattice edges $E$ satisfies
Euler's relation $E=N+N_D-2$ and each lattice edge contributes a
factor $mn$. Eq.~(\ref{psdual}) then shows that the powers of $mn$ cancel
after a pair of duality transformations.

With the notation
\begin{equation}
x_i \equiv e^{K_i}/e^{K_0},\mbox{\hspace{2mm}}
x_i^* \equiv e^{K^*_i}/e^{K^*_0},\mbox{\hspace{2mm}}
(i=1,2,3)
\end{equation}
the Boltzmann factor can be written as
\begin{displaymath}
B(\sigma,\tau|\sigma',\tau') =
\mbox{\hspace{54mm}}
\end{displaymath}
\vspace{-5mm}
\begin{equation}
\mbox{\hspace{4mm}}
e^{K_0}
x_1^{(1-\delta_{\sigma \sigma'}) \delta_{\tau \tau'}} 
x_2^{\delta_{\sigma \sigma'}(1- \delta_{\tau \tau'})}
x_3^{(1-\delta_{\sigma \sigma'})(1- \delta_{\tau \tau'})} \,.
\label{bfact}
\end{equation}

Phase transitions will naturally occur in the $\mathbf{x}=(x_1,x_2,x_3)$
parameter space. Thus, the free energy of the $(m,n)$ model is expected
to be nonanalytic on a corresponding surface $\Sigma$. Generally, as the
temperature variable in a given model varies, the point $(x_1,x_2,x_3)$
traces out a certain "thermodynamic" path $\Gamma$ in $\mathbf{x}$, and
the model exhibits a transition whenever $\Gamma$ crosses $\Sigma$.

Using (\ref{dual}) we find the transformation
\begin{eqnarray}
x_1^*&=&\frac{1}{\Delta}[1-x_1+(n-1)x_2-(n-1)x_3] \nonumber\\
x_2^*&=&\frac{1}{\Delta}[1-(m-1)x_1-x_2-(m-1)x_3] \nonumber\\ 
x_3^*&=&\frac{1}{\Delta}[1-x_1-x_2+x_3]            
\label{dual2}
\end{eqnarray}
where
\begin{equation}
\Delta=1+(m-1)x_1+(n-1)x_2+(m-1)(n-1)x_3 \,.
\end{equation}
The square lattice maps onto itself under the dual transformation, so
that the free energies at two points ${x_i}$ and ${x_i^*}$ satisfying  
Eq.~(\ref{dual2}) are related.  One also verifies that the subspace
\begin{equation}
\Delta=\sqrt{mn}
\end{equation}
is self-dual under the transformation Eq.~(\ref{dual2}) (but a point
in this plane does, in general, not map on itself).

We note that the partition sum of the cubic loop model described 
by Eq.~(\ref{Zcub2}) with $v=0$, i.e.,
the model of Eq.~(\ref{Zcub4}), can be written as 
\begin{equation}
Z_{\rm cub}=   \sum_{\{s\},\{\tau\}} \prod_{\langle{ij}\rangle}
[1+nxs_is_j\delta_{\tau_i\tau_j}] \,.
\end{equation}

By introducing $\sigma_k\equiv (s_k+3)/2$, and the identities
\begin{equation}
s_i s_j=2\delta_{\sigma_i\sigma_j}-1
\end{equation}
we find the equivalence
\begin{equation}
Z_{\rm cub}=Z_{(2,n)}(e^{K_0},e^{K_1},e^{K_2},e^{K_3})
\end{equation}
with
\begin{equation}
e^{K_0}=1+nx \,, \mbox{\hspace{2mm}} e^{K_1}=1-nx \,,
\mbox{\hspace{2mm}} e^{K_2}=e^{K_3}=1
\end{equation}
or
\begin{equation}
x_1=\frac{1-nx}{1+nx} \,,\mbox{\hspace{2mm}} x_2=\frac{1}{1+nx} \,,
\mbox{\hspace{2mm}} x_3=\frac{1}{1+nx} \,.
\end{equation}
This specifies the mapping of the $n$-component cubic loop model on 
the $(N_\alpha=2,N_\beta=n)$ model.
Using the dual transformation (\ref{dual2}), this gives rise to
\begin{equation}
x_1^*=x_3^*=x, \mbox{\hspace{5mm}} x_2^*=0 \,.
\end{equation}
Hence the dual thermodynamic path $\Gamma$ of the cubic loop model 
is the straight line connecting $(0,0,0)$ and $(1,0,1)$, a result
valid for all $n$.

For two special cases, namely $n=1$ and $n=2$, the critical point
of the cubic loop model sits at the intersection of the critical 
surface $\Sigma$ and the thermodynamic path $\Gamma$ in the $x_2=0$
plane.

We first consider the case $n=1$, or $\delta_{\tau_i\tau_j}=1$, in
which the model simply reduces to the Ising model. The Boltzmann
factor (\ref{bfact}) assumes the form
\begin{equation}
B(\sigma,\tau|\sigma',\tau')=e^{K_0}x_1^{1-\delta_{\sigma\sigma'}} 
\end{equation}
for any values of $x_2, x_3$. 
The critical surface is thus  $x_1=\sqrt{2}-1$.
This plane intersects the thermodynamic path $\Gamma$ at 
$x_1=x_3=x=\sqrt{2}-1$, so that the critical point of the cubic loop 
model is $x_c=\sqrt{2}-1$. 

For the case $n=2$, the model, i.e., the $(2,2)$ model, is the well-known
Ashkin-Teller model \cite{AT}.  The shape and location of $\Sigma$ have
been discussed by Wu and Lin \cite{WL}.  The thermodynamic path $\Gamma$ 
crosses the critical surface $\Sigma$ at $x_1=x_3=1/2$ within the 
$x_2=0$ plane \cite{WL}. Thus the critical point of the cubic loop
model occurs  at $x_c=1/2$.

While it is known that most of the self-dual plane of the
($N_\alpha,N_\beta$) model is noncritical, it is interesting that for
both $n=1$ and $n=2$, the critical points of the cubic loop model
are actually located in the self-dual plane. If this is true for
general $1<n<2$, we would have
\begin{equation}
1+x_1+(n-1)x_3=\sqrt{2n}
\label{conjd}
\end{equation}
on the $x_2=0$ plane, and $x_1=x_3=x$, because the critical point lies on 
the thermodynamic path $\Gamma$. Thus the critical value of $x$ would be 
\begin{equation}
x_c(n)=(\sqrt{2n}-1)/n {\mbox{\rm ~~ for ~~}}  1\leq n \leq 2 \, .
\label{conj}
\end{equation}
This possibility will be investigated numerically in
Sec. \ref{crline}.
\section {The transfer matrix }
\label{trmat}
The transfer-matrix method used here is related to that used in 
Ref.~\onlinecite{BNcub}, and it uses in addition some of the techniques
described
in Refs.~\onlinecite{BNp} and \onlinecite{BN} for the random-cluster
and the O($n$) model respectively. The full description of the transfer
matrix is somewhat elaborate, and here we only provide a general
outline, supplemented with more detailed information where the procedure
is different from those in the references given.

As in Ref.~\onlinecite{BNcub}, the transfer matrix is constructed on
the basis of a graph representation of the cubic model that allows
$n$ to be non-integer. However, the present work is restricted
to the case $v=0$, so that the graphs are restricted to be even. This
allows the use, given a system size, of a smaller transfer matrix than
that used in Ref.~\onlinecite{BNcub}.
We define the model on an $L \times m$ lattice ${\cal L}_m$ wrapped
on a cylinder, such that the finite-size parameter $L$ is the
circumference of the cylinder. The definition of the transfer matrix
can be illustrated by appending row $m+1$ and determining how this
affects the partition sum of the model. The lattice ${\cal L}_m$ has
an open end at row $m$; there are $L$ `dangling' edges that will serve
to connect to row $m+1$ later. Whereas the partition sum Eq.~(\ref{Zcub3})
allows only for closed loops, the bond configuration on ${\cal L}_m$
may be considered a part of a larger graph so that we allow the
dangling edges to be occupied by loop segments. But all sites are
still restricted to connect to an even number of bonds.
For the construction of the transfer matrix, we need a coding of all
possible ways (called connectivities) that the dangling bonds can
be connected by the graph $\{b\}$ on ${\cal L}_m$. This coding assigns
a unique integer $1,2,\cdots$, which will serve as the transfer-matrix
index, to each connectivity. Some of the dangling
edges may be empty, i.e., $b_{ij}=0$. The remaining dangling edges,
i.e., the dangling bonds, form
a `dense connectivity' without vacant positions. Note that these dense
connectivities satisfy a `well-nestedness' principle which asserts that,
if positions $i$ and $k$ are connected, and $j$ and $l$ are connected, 
with $i<j<k<l$, then all $i,j,k,l$ must be connected. Thus, these dense
connectivities form a subset of the `random cluster' or `Whitney'
connectivities defined in Ref.~\onlinecite{BNp}. The number of dangling
bonds that are connected by a path of bonds must always be even for
the cubic model; this restriction does not apply to those of
Ref.~\onlinecite{BNp}. By simply excluding the odd connectivities, we
thus find coding and decoding algorithms for the dense cubic
connectivities. The coding of the general cubic connectivities
including vacant positions  then follows analogous to the case
of the `magnetic' connectivities in Ref.~\onlinecite{BNp}.

This coding allows one to divide the partition sum $Z^{(m)}$ of the
$m$-row system in contributions
corresponding with different connectivities:
\begin{equation}
 Z^{(m)}= \sum_\alpha  Z^{(m)}_\alpha \,.
\label{Z-sum}
\end{equation}
Next, we append a new row $l_{m+1}$ to the lattice:
${\cal L}_{m+1} \equiv {\cal L}_{m} \cup l_{m+1}$, and express the
restricted partition sums $Z^{(m+1)}_\alpha$ as a linear combination
of the $Z^{(m)}_\beta$. This is possible because the weight due to
the newly appended row is completely determined by the bond
variables connecting to the appended vertices and the `old'
connectivity $\beta$, and this information also determines the `new'
connectivity $\alpha$. We use $b_{m+1}$ to denote the $2L$ appended
bond variables, and $\mu(\beta,b_{m+1})$ to denote the function that
determines $\alpha$.
The weight factor associated with the new row satisfies 
\begin{equation}
w(\beta,b_{m+1}) =(2n)^L x^{\Delta n_b} n^{\Delta n_l} \,,
\label{wfac}
\end{equation}
where $\Delta n_b$ is the number of appended bonds and
$\Delta n_l$ is the number of loops closed by these bonds.
The recursion connecting the restricted sums is
\begin{equation}
 Z^{(M+1)}_\alpha = \sum_{\beta} T_{\alpha \beta} Z^{(M)}_{\beta} \,,
\label{Z-recursion}
\end{equation}
in which the transfer matrix \( T_{\alpha \beta} \) is defined by
\begin{equation}
  T_{\alpha \beta} =(2n)^L  \sum_{b_{m+1}}
\delta_{\alpha,\mu(\beta,b_{m+1})} x^{\Delta n_b} n^{\Delta n_l} \,.
\label{T-def}
\end{equation}
In actual calculations, the transfer matrix is represented as 
the product of $L$ sparse matrices, each of which appends one 
new vertex of the $(m+1)$-th row. The first vertex of a new row
increases the number of dangling edges to $L+2$, so that the sparse
matrices assume a larger size than $T_{\alpha \beta}$.
After appending the last vertex of that row, the number of dangling
edges decreases  to $L$.  This technical point was described
in some detail for the related case of the O($n$) model on the
square lattice \cite{BN}.

The sparse-matrix technique makes it unnecessary to store the full
transfer matrix ${\bf T}$. Some of its eigenvalues can be obtained by
repeated multiplication of a vector by ${\bf T}$, and analysis of the
resulting vector sequence.  Since $T_{\alpha \beta}$ is not symmetric
in general, we used the method of projection to a Hessenberg matrix 
as described in Ref.~\onlinecite{BNp}. We restricted the calculations to
vectors with `translation symmetry', i.e., vectors that are invariant 
under a permutation of connectivities corresponding with a cyclic
permutation of the dangling edges.
In general, the largest eigenvalue $ \Lambda^{(0)}_L$ determines the
free energy $f(L)$ per site in the limit of an infinitely long cylinder
($ m \rightarrow \infty$):
\begin{equation}
 f(L) = L^{-1}  \ln \Lambda^{(0)}_L \,.
\label{flambda}
\end{equation}
Furthermore, the next largest eigenvalues $\Lambda^{(i)}_L$
($i=1,2,\cdots)$ determine the correlation lengths $\xi_i(L)$ of
various types of correlation functions. The latter types depend
on the symmetry of the corresponding eigenvector and on possible 
modifications of ${\bf T}$.
In particular, the correlation length $\xi_t(L)$ of the energy-energy
correlation function is determined by the gap between the two largest
eigenvalues:
\begin{equation}
\xi_t^{-1}(L) = \ln (\Lambda^{(0)}_L/\Lambda^{(1)}_L) \,.
\end{equation}

For the cubic model, magnetic correlations can be represented, in analogy
with the O($n$) model, by graphs with odd vertices on the correlated
sites. For the present model that means sites connected to one or three
bonds. The two correlated sites, which are 
placed far apart in the length direction of the cylinder, must be
connected by the graph, i.e., belong to the same component of the graph.
This additional component does not follow the rules of `evenness'
listed earlier. The number of dangling bonds at the open end of the
cylinder (between the correlated sites) connecting to the additional
odd vertex must be odd. To describe such `magnetic' graphs we define
a new set of connectivities in which one group of connected dangling
bonds is odd. This leads to a modified transfer matrix, which may
alternatively be interpreted as the magnetic sector of a larger transfer
matrix whose vector space includes both even and odd connectivities. 
The gap between $\Lambda^{(0)}_L$ and the largest eigenvalue
$\Lambda^{(2)}_L$ in the magnetic sector 
determines the magnetic correlation length $\xi_h(L)$ as
\begin{equation}
\xi_h^{-1}(L) = \ln (\Lambda^{(0)}_L/\Lambda^{(2)}_L) \,.
\end{equation}

A different type of magnetic gap is associated with the density of
the loops spanning the circumference of the cylinder. The weight of
these loops is modified by assigning a bond weight $-x$ to one bond
in each row. A loop spanning the circumference contains an odd number
of these modified bonds, and its weight thus changes sign.
All other loops contain an even number of modified bonds and their
weights are thus unchanged. We denote the largest eigenvalue of 
the modified transfer matrix by $\Lambda^{(3)}_L$. It determines the
length scale $\xi_m(L)$ that may be associated with the effect of an
`antiferromagnetic seam' running along the cylinder.
For the critical Ising case $n=1$, both magnetic length scales are
related by duality, but this is not so for general $n$.
The corresponding length scale is given by
\begin{equation}
\xi_m^{-1}(L) = \ln (\Lambda^{(0)}_L/\Lambda^{(3)}_L) \,.
\end{equation}
In the actual transfer-matrix calculations, we have used finite sizes
up to $L=15$ in the nonmagnetic sector, which then has dimensionality
2\,004\,032, and up to $L=14$ in the magnetic sector, which then has
dimensionality 3\,856\,582.

\section {Determination of the critical line of the cubic model}
\label{crline}
The asymptotic behavior of the magnetic correlation length $\xi_h(L)$
near a critical point can be expressed in terms of the scaled gap
\begin{equation}
X_h(t,u,L) \equiv \frac {L}{2\pi \xi_h(t,u,L)} \,,
\label{scg}
\end{equation}
where $t$ parametrizes the distance to the critical point, and $u$ 
represents an irrelevant field. Renormalization arguments \cite{Suz},
scaling \cite{FSS}, and conformal invariance \cite{Cardy} predict that
for large $L$ 
\begin{equation}
X_h(t,u,L) \simeq X_h + a_1 L^{y_t} t + b_1 L^{y_u} u + \cdots \,,
\label{scxh}
\end{equation}
where  $X_h$ is the magnetic scaling dimension, $y_t$ the temperature
exponent, and $y_u$ the exponent of the field $u$, and $a_1$ and $b_1$  
are unknown amplitudes. Further corrections may also be present.
Since we have an algorithm available that calculates $X_h(t,u,L)$
(with $t$ and $u$ expressed as a function of $x$), we can
estimate the critical point by numerically solving $x$ in the equation
\begin{equation}
X_h(x,L) = X_h(x,L+1) \,,
\label{xsol}
\end{equation}
which is a form of `phenomenological renormalization' \cite{PhR}.
After substitution of Eq.~(\ref{scxh}) one finds that, at the solution,
$t$ and $u$ satisfy
\begin{equation}
t \propto u L^{y_u-y_t}  \,.
\label{tdev}
\end{equation}
Since $y_t>0$ and $y_u<0$, we expect that $t\to 0$ for $L \to \infty$,
i.e., the solutions of Eq.~(\ref{xsol}), which we denote $x^{(0)}(L)$,
converge to the critical point. These solutions 
were fitted by solving for $x^{(1)}(L)$, $c^{(1)}(L)$ and $y_u-y_t$
in the three following equations with $L'=L$, $L-1$, and $L-2$:
\begin{equation}
x^{(0)}(L') = x^{(1)}(L) +c^{(1)}(L) L'^{y_u-y_t} \,,
\label{xfit}
\end{equation} 
which leads to a sequence $x^{(1)}(L)$ that is shorter than the
original sequence $x^{(0)}(L)$ but usually shows faster apparent
convergence.
Another iteration step can be attempted on the basis of
\begin{equation}
x^{(1)}(L') = x^{(2)}(L) +c^{(2)}(L) L'^{y_{u}'-y_t}\,,
\label{xfit2}
\end{equation}
which may lead to even better estimates of the critical point.

A similar analysis of the critical point can be performed on
the basis of the scaled interface gap
\begin{equation}
X_m(t,u,L) \equiv \frac {L}{2\pi \xi_m(t,u,L)} \, ,
\label{scig}
\end{equation}
using the same type of fits as for the scaled magnetic gap. 

We have also attempted to find solutions of Eq.~(\ref{xsol}) with
$\xi_h$ replaced by the energy-energy correlation length, but here
complications arise. The functions $X_t(x,L)$ typically display an
extremum near the critical point, and solutions of the scaling equation
Eq.~(\ref{xsol}), with $X_t$ instead of $X_h$, do not always exist.
In particular for $n>1$ we did not obtain a satisfactory
set of solutions, and we have not pursued this way to obtain further
data for the critical point. Instead, we located the extremum of
$X_t(x,L)$ as a function of $x$. The finite-size-scaling equation for
the correlation length indicates that 
this extremum will converge to the critical point. 

\begin{figure}
\includegraphics{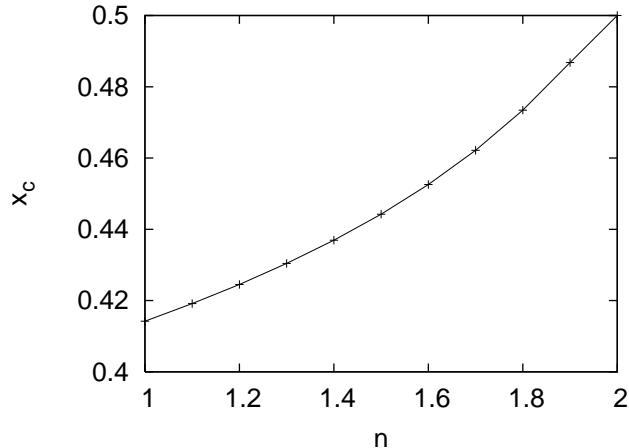}
\caption{Critical points of the $v=0$ cubic model. The data points show
the numerical results of the transfer-matrix analysis, and the curve
is added as a guide to the eye. The estimated error bars are smaller
than the size of the symbols.
}
\label{fig1}
\end{figure}

The estimated critical points are shown in Fig.~\ref{fig1}.
For $1 < n < 2$ they do not agree with Eq.~(\ref{conj}). For instance,
that equation would predict $x_c=0.48803\cdots$ for $n=1.5$, which is
incompatible with the numerical result (see also Table \ref{tab_1}).
Thus we have to conclude that Eq.~(\ref{conj}), which does indeed lack
a solid basis, is not valid for all values of $n$.

Several modified fitting procedures were applied. Assuming that
the cubic model reduces to the O($n$) universality class, we have
analytic evidence for the values of $y_u$ and $y_{u}'$ as a function
of $n$. First, according to the renormalization scenario, the cubic
perturbation of the O($n$) symmetry corresponds with an irrelevant
field with an exponent \cite{NienhuisD&L}
\begin{equation}
y_c = \frac{(1-g)(1+3g)}{2g} \,,
\label{xcp}
\end{equation}
where $\cos(\pi g)=-n/2$ and $1 \leq g\leq2$. Corrections due to
this field are expected to be serious for $n \to 2$, since $y_c \to 0$.
Second, irrelevant temperature-like fields may correspond with some
scalar operators whose dimensions are entries $X_{1,q}$ with $q>3$
in the Kac table \cite{FQS,Kac}
\begin{equation}
X_{p,q}=\frac{[p(m+1)-qm)]^2-1}{2m(m+1)} 
\label{kac}
\end{equation}
with $m=1/(g-1)$ for the cubic model \cite{BCN}.
For $q=5$ we find an irrelevant exponent
\begin{equation}
y_{i}=\frac{6 g -12}{g} \, .
\label{q5}
\end{equation}
which has small values for $n>0$ but becomes marginal when
$n \to -2$.
The final results, and their estimated errors, were obtained from the
analyses of the three types of scaled gap, and the degree
of consistency between different types of fits mentioned above. The best
estimates obtained from $X_h(t,u,L)$ and $X_m(t,u,L)$, and the overall
best estimates which also include data from $X_t(t,u,L)$, are shown in
Table \ref{tab_1}.

\begin{table}
\caption{Critical points $x_c$ as determined from the scaling formulas
for the magnetic and interface correlation length for system sizes $L$
and $L+1$, and from the extrema of the energy-like correlation length
as $x$ is varied.  The estimated numerical uncertainty in the last
decimal place is shown in parentheses. The `best estimates' are based
on the results in the two preceding columns and on an analysis of the
minima in the functions $X_t(x,L)$ as described in the text. For $n=1$
we find accurate agreement with the exact result $x_c=\sqrt{2}-1$, and
for $n=2$ with $x_c=1/2$.  }
\label{tab_1}
\begin{center}
\begin{tabular}{||l|l|l|l||}
\hline
$n$ & $x_c$ (from $X_h$) & $x_c$ (from $X_m$) & $x_c$ (best estimates) \\   
\hline
1.0 &    0.4142134 (2)   & 0.4142135 (1) &    0.4142135 (1) \\
1.1 &    0.419155  (5)   & 0.419154  (2) &    0.419154  (2) \\
1.2 &    0.424530  (5)   & 0.424527  (2) &    0.424528  (2) \\
1.3 &    0.43042   (1)   & 0.430415  (5) &    0.430416  (5) \\
1.4 &    0.43695   (2)   & 0.436935  (5) &    0.43694   (1) \\
1.5 &    0.44423   (5)   & 0.44424   (1) &    0.44424   (1) \\
1.6 &    0.45257   (5)   & 0.45254   (1) &    0.45254   (1) \\
1.7 &    0.46212   (5)   & 0.46214   (1) &    0.46214   (1) \\
1.8 &    0.4736    (1)   & 0.47345   (2) &    0.47346   (2) \\
1.9 &    0.4869    (2)   & 0.48673   (5) &    0.48675   (5) \\
2.0 &    0.5000000 (1)   & 0.4999999 (1) &    0.5000000 (1) \\
\hline
\end{tabular}
\end{center}
\end{table}

\section {Universal properties of the cubic model }
\label{uniprop}
The asymptotic finite-size dependence of the free energy per site at
the critical point is
\cite{BCN,Affl}
\begin{equation}
 f(L) \simeq f(\infty) + \frac{\pi c}{ 6L^2} \,,
\label{scf}
\end{equation}
where $c$ is the conformal anomaly of the model, which characterizes
universality classes and determines sets of critical
exponents \cite{CardyD&L,BPZ}.
We have calculated the finite size data for the free energy at the
extrapolated critical points, and estimated $c$ as $c^{(1)}(L)$
from the free energy density for two consecutive system sizes 
by solving
\begin{equation}
c^{(1)}(L) =6 [f(L) -f(L+1)]/ [ \pi \{ 1/L^2- 1/(L+1)^2 \} ] \,.
\label{ces}
\end{equation}
This leads to a sequence of estimates of $c$ that can be extrapolated by
means of power-law fits, analogous to the procedure used to determine the
critical points. After a second iteration step, the estimates of $c$
seem almost converged, in the sense that the results for the largest few
values of $L$ display differences of only a few times $10^{-5}$. But at
the same time these data display a shallow extremum (except for $n=1$ and
2, where the apparent convergence is much better), so that it is difficult
to estimate the uncertainty in the extrapolated results.
Because the first iteration step shows that the  finite-size dependence of
the estimates of $c$ is close to $-2$, we have also applied iteration steps
with the exponent fixed at this value. The results were similar to
those obtained with free exponents, and again displayed a shallow extremum.
Under these circumstances we made a crude error estimate of ten times the 
difference between the two estimates for the largest available $L$ values,
after two iteration steps. The best estimate was taken by extrapolating
the last two estimates, using again ten times the aforementioned difference.
A better apparent convergence was found when a fixed exponent $y_c -2$
was used in the second iteration step. The results are shown in Table
\ref{tab_2}. The numerical errors were estimated from the finite-size
dependence of the results of the last fit procedure, except for $n=1.9$,
where the error bars of both fit procedures did not overlap and we used
the difference between both types of fit instead.
\begin{table}
\caption{ Conformal anomaly and temperature scaling dimension $X_t$ as 
determined by transfer-matrix calculations described in the text. Estimated 
error margins in the last decimal place are given in parentheses. 
For comparison, we include the theoretical values of the O($n$) model 
for $n<2$, and of the Ashkin-Teller model for $n=2$.  The numerical results 
are indicated by `num', theoretical values by `theory'.}
\label{tab_2}
\begin{center}
\begin{tabular}{||l|l|l|l|l||}
\hline
$n$&$c({\rm num})$&$c({\rm theory})$&$X_t({\rm num})$&$X_t({\rm theory})$\\
\hline
1.0 & 0.50000 (1)  & 0.50000 & 1.000000 (1) & 1.00000 \\
1.1 & 0.54820 (2)  & 0.54820 & 1.0428   (5) & 1.04269 \\
1.2 & 0.59640 (2)  & 0.59639 & 1.0890   (5) & 1.08840 \\
1.3 & 0.64465 (5)  & 0.64465 & 1.1385   (5) & 1.13782 \\
1.4 & 0.6931  (1)  & 0.69309 & 1.192    (1) & 1.19187 \\
1.5 & 0.7418  (1)  & 0.74184 & 1.251    (2) & 1.25189 \\
1.6 & 0.7912  (1)  & 0.79106 & 1.319    (2) & 1.31996 \\
1.7 & 0.8410  (1)  & 0.84096 & 1.405    (5) & 1.39962 \\
1.8 & 0.8920  (2)  & 0.89186 & 1.49     (2) & 1.49783 \\
1.9 & 0.9464  (9)  & 0.94432 & 1.60     (5) & 1.63279 \\
2.0 & 1.00000 (1)  & 1       & 1.500000 (1) & 3/2     \\
\hline
\end{tabular}
\end{center}
\end{table}
Most of our results are in good agreement with the theoretical values:
\begin{equation}
c=1-\frac{6 (g-1)^2}{g} \,.
\end{equation}
which follow after the substitution of the formula \cite{BCN}
$m=1/(g-1)$ in the relation \cite{CardyD&L} between $m$ and $c$, i.e.,
$c=1-6/[m(m+1)]$. But the result for $n=1.9$ does not
agree well with the theoretical value; we note that the small value of 
the cubic crossover exponent, which becomes marginal at $n=2$, may well
lead to imprecise results and error estimates.

Next, we analyze the results for the magnetic gaps.
After substitution of the solutions of Eq.~(\ref{xsol}), which behave
as Eq.~(\ref{tdev}), in Eq.~(\ref{scxh}), one finds that the magnetic 
scaled gaps at the solutions converge to $X_h$ as:
\begin{equation}
X_h(L)=X_h+r u L^{y_u}+\cdots \,,
\end{equation} 
where $r$ is an unknown constant.
The magnetic scaled gaps at the solutions of the scaling equation in the
previous section were already available. They were fitted using a similar
procedure as used for the determination of the critical points.
We found that the leading irrelevant exponent was consistent with the
predicted cubic perturbation exponent given in Eq.~(\ref{xcp}), and we
accordingly treated $y_u$ as a known parameter in the fits.
The extrapolated magnetic scaling dimensions are shown in Table
\ref{tab_3}.
\begin{table}
\caption{
The magnetic dimension $X_h$ and the interface dimension $X_m$
as extrapolated from their values 
at the solutions of the scaling equation of the correlation length.
Estimated numerical uncertainty in the last decimal place are given
in parentheses. For comparison, we include the theoretical values of 
the O($n$) model for $n<2$, and of the Ashkin-Teller model for $n=2$.
The numerical results are indicated by '(num)', the theoretical
values by '(theory)'. }
\label{tab_3}
\begin{center}
\begin{tabular}{||l|l|l|l|l||}
\hline
$n$&$X_h({\rm num})$&$X_h$({\rm theory})&$X_m({\rm num})$&$X_m$({\rm theory})\\
\hline
1.0 & 0.12499 (1) & 0.12500 & 0.12499  (1) & 1/8       \\
1.1 & 0.12647 (1) & 0.12668 & 0.14105  (5) & 0.14101   \\
1.2 & 0.1280  (1) & 0.12826 & 0.15814  (5) & 0.15815   \\
1.3 & 0.1296  (1) & 0.12973 & 0.1767   (1) & 0.17668   \\
1.4 & 0.1307  (1) & 0.13107 & 0.1969   (2) & 0.19695   \\
1.5 & 0.1316  (1) & 0.13224 & 0.2195   (5) & 0.21946   \\
1.6 & 0.1320  (1) & 0.13319 & 0.244    (1) & 0.24499   \\
1.7 & 0.1335  (3) & 0.13382 & 0.274    (2) & 0.27486   \\
1.8 & 0.1308  (2) & 0.13393 & 0.313    (2) & 0.31169   \\
1.9 & 0.128   (1) & 0.13300 & 0.350    (5) & 0.36230   \\
2.0 & 0.12498 (1) & 1/8     & 0.37500  (1) & 3/8       \\
\hline
\end{tabular}
\end{center}
\end{table}
Again, different fit procedures, with the finite-size exponent left
free, and more iteration steps, were applied. The error estimates are
based on the apparent convergence and on the differences between the
various types of fit.  The final results appear to agree with the
theoretical values for the O($n$) universality class \cite{N}:
\begin{equation}
X_h=\frac{g}{8}-\frac{(1-g)^2}{2 g} \,.
\end{equation}
where we note that the Ashkin-Teller model (or $(2,2)$ model) has the
same magnetic scaling dimension \cite{E,dN} $X_h=1/8$ as the $O(2)$ model.

A similar analysis was performed on the scaled interface gaps at the 
solutions of the scaling equation for interface scaled gap. These gaps 
converge to the interface scaling dimension. The results for the
interface scaling dimension are shown in Table \ref{tab_3}. These 
results are to be compared with the known interface exponent in the
O($n$) universality class \cite{BN}, which are
given by the entry $p=1,q=2$ in the Kac table:
\begin{equation}
X_m=\frac{3}{2g}-1 \,.
\label{xm}
\end{equation}
which is obtained by the substitution of $m=1/(g-1)$ in the more
common form $X_{1,2}=[(m-1)^2-1]/2 m(m+1)$ of the Kac formula.

Our result for $X_m$ at $n=2$ is different from $X_{1,2}=1/2$ and thus
illustrates that the $n=2$ cubic loop model falls outside the O(2)
universality class.  It is 
to be compared with results for the Ashkin-Teller model \cite{Kn,dN}
which are also summarized by Baxter \cite{Baxter}. In the notation
used there, we have the exact result $\beta^{8V}_e=3/4$ for the end point
of the Ashkin-Teller line at $K \to \infty$. This exponent may, in our
notation, be put equal to $X_m/(2-X_t)$ where $X_t=3/2$. This does indeed 
lead to $X_m=3/8$.

We also calculated the temperature scaled gaps $X_t(x,L)$ at the
extrapolated critical points. 
We expect the following behavior of these gaps:
\begin{equation}
X_t(L)=X_t+ p u L^{y_u}+\cdots \,,
\end{equation}
where $p$ is another unknown constant.
Similar fits as before lead to results for the temperature scaling 
dimension that are included in Table \ref{tab_2}. The results agree with
the theoretical prediction \cite{N} for the O($n$) model
\begin{equation}
X_t=\frac{4}{g}-2 \,,
\end{equation}
except for the case $n=2$, where our numerical result goes to $3/2$
in accordance with the exact result \cite{Kn,dN} for the temperature
scaling dimension of the Ashkin-Teller model for $K \to \infty$.

\section {Discussion}
\label{disc}
In general, the discreteness of the $n$-component cubic model defined
by Eq.~(\ref{Hcub}) will enforce the existence of a long-range ordered
phase at sufficiently low temperatures, also for $n>2$. However, our
transfer-matrix calculations for $v=0$ in Eq.~(\ref{specc}) did not
yield any evidence for a phase transition to the ordered phase for $n>2$.
The absence of a phase transition is understandable in terms of the 
parameters $K$ and $M$ in Eq.~(\ref{Hcub}), because Eq.~(\ref{specc})
restricts these parameters to $K+M \leq \ln 2$. This quantity, which
specifies the energy difference between parallel 
($\vec{s_i} \cdot \vec{s_j}=1$) spins and perpendicular
($\vec{s_i} \cdot \vec{s_j}=0$) spins, is not sufficient to reach a
long-range-ordered phase for $n>2$. The maximum value $K+M=\ln 2$
along the critical line is reached for $n=2$ where $K=\infty$.

For $n\leq 2$ we find clear evidence for a phase transition to the 
ordered phase. In the interval $1 \leq n \leq 2$ we have determined 
the critical points of the $n$-component cubic model as given by
Eqs.~(\ref{Hcub}) and (\ref{specc}), or by Eq.~(\ref{Zcub4}).
For the exactly solved cases $n=1$ and $n=2$ we find good agreement
with the exact values as given in Table \ref{tab_1}.
Our results for the scaling dimensions, i.e., $X_t$ associated 
with  the temperature, $X_h$ associated with the magnetic field, and
$X_m$ associated with the introduction of an antiferromagnetic `seam'
in the model, agree accurately with the O($n$) universality classes,
with the exception of the case $n=2$ where the model reduces to a
special case of the Ashkin-Teller model. For the latter case $n=2$, our
results for the scaling dimensions agree with the exact results for the
Ashkin-Teller model. The fact that these scaling dimensions are different
from those for the O(2) universality class is related with the cubic
anisotropy which may be seen as a perturbation of the O($n$) symmetry. 
This perturbation is irrelevant for $n<2$ but marginal \cite{NienhuisD&L}
for $n=2$. This proof of irrelevance for $n<2$ applies to
small cubic perturbations of the isotropy. Our numerical results show
that the cubic perturbation remains irrelevant even in the extreme
anisotropic case described by Eqs.~(\ref{Hcub}) and (\ref{specc}).
However, when $n$ approaches 2, the exponent $y_c$ associated with the
cubic anisotropy field approaches marginality and our numerical results
thus become less accurate. The cubic anisotropy is truly marginal at
$n=2$ and parametrizes the Ashkin-Teller model. For this reason, the
cubic crossover phenomena are absent for $n=2$ and the numerical results
are again relatively accurate.
The temperature scaling dimension of the cubic model and its consistency
with O($n$) universality were already determined \cite{BNcub} for a
range $n \leq 1$. The present results extend the range of $n$ and provide
additional evidence concerning the dimensions $X_h$ and $X_m$.
The numerical accuracy of our analysis is such that the O($n$)
universality of the cubic model seems reasonably convincing. We note
that numerical results for the O($n$) model on the square lattice
\cite{BN} are even more accurate; however, for that case the exact
critical point is known, and the cubic anisotropy field, and thereby
the leading corrections to scaling, vanish.

\acknowledgments
This research is supported in part by the Dutch FOM
foundation (`Stichting voor Fundamenteel Onderzoek der Materie') which
is financially supported by the NWO (`Nederlandse Organisatie voor
Wetenschappelijk Onderzoek'), by the  National Science Foundation of
China  under Grant \#10105001, and by a grant from Beijing Normal
University.

                                                       
\end{document}